\documentclass[aps,onecolumn,12pt,superscriptaddress,showpacs]{revtex4-1}

\usepackage{graphicx,color}
\usepackage{amsmath}
\usepackage{amssymb}
\usepackage{subfigure}
\newcommand{\be}{\begin{equation}}
\newcommand{\ee}{\end{equation}}

\begin{document}

\title{Angular momentum blockade in nanoscale high-$T_c$ superconducting grains}

\author{Francesco Mancarella}
\affiliation{Nordic Institute for Theoretical Physics (NORDITA), Roslagstullsbacken 23, S-106 91 Stockholm, Sweden}
\affiliation{Department of Theoretical Physics, KTH Royal Institute of Technology, 
SE-106 91 Stockholm, Sweden}

\author{Alexander V. Balatsky}
\affiliation{Nordic Institute for Theoretical Physics (NORDITA), Roslagstullsbacken 23, S-106 91 Stockholm, Sweden}
\affiliation{Theoretical Division, Center for Integrated Nanotechnologies, Los Alamos National Laboratory, Los Alamos, NM 87545, USA.}

\author{Mats Wallin}
\affiliation{Department of Theoretical Physics, KTH Royal Institute of Technology, 
SE-106 91 Stockholm, Sweden}

\author{Anders Rosengren}
\affiliation{Department of Theoretical Physics, KTH Royal Institute of Technology, 
SE-106 91 Stockholm, Sweden}

\begin{abstract}
We discuss the angular momentum blockade in small $d$-wave superconducting grains in an external field. We find that abrupt changes in angular momentum state of the condensate, {\em  angular momentum blockade},  occur as a result of changes in the angular momentum of the condensate in an external magnetic field.
 The effect represents a  direct  analogy with  the Coulomb blockade. We use the Ginzburg-Landau  formalism to illustrate how a magnetic field induces a deviation from the $d$-wave symmetry which is described by a ($d_{x^2-y^2}+id_{xy}$)-order parameter.
We derive the behavior of the volume magnetic susceptibility as a function of the magnetic field, and corresponding magnetization jumps at critical values of the field that should be experimentally observable in superconducting grains.

\end{abstract}

\maketitle

%\begin{small}\textbf{Keywords:}\end{small} \begin{footnotesize} ..... \end{footnotesize}\\

\begin{small}\textbf{PACS number:}\end{small} 74.20.Rp, 	 74.25.-q,  74.78.Na \\

\begin{small}\textbf{Contact information}

Corresponding author: Francesco Mancarella; e-mail: framan@kth.se ; Fax: +46 8 5537 8404; Address: NORDITA, Roslagstullsbacken 23, 106 91 Stockholm, Sweden \end{small}\\

\newpage

\section{Introduction}
The precise nature of the superconducting state in cuprate superconductors has been discussed extensively since the discovery of high-$T_c$ superconductivity \cite{Bednorz86}. While most of the data can be well covered assuming a pure $d$-wave symmetry of the order parameter \cite{Wollman93, Hardy93, Shen93, Tsuei98}, the identification of the  precise symmetry of the order parameter  remains one of the active areas of research. 
One can imagine that the pairing state symmetry is affected by the crystal field and, as is the case of YBCO, by the presence of oxygen chains \cite{Xiang96, Muller02}. 
Moreover, even if the pairing state symmetry is a simple $d$-wave, it can be modified and distorted by application of an external field \cite{Laughlin98} and by scattering off defects \cite{Balatsky95, Movshovich98, Balatsky06}. These distortions can also depend on the doping and the nature of correlation effects in these materials \cite{Muller95}. One might for example expect that the symmetry of the superconducting order changes as a function of doping and therefore this pairing symmetry contains useful information about the microscopic interactions responsible for pairing. 
Indeed recent results suggest that nanoscale $d$-wave superconductors can be fully gapped and this minimal gap (on the scale of 10 mK) can be modified by an external magnetic field \cite{Gustafsson13}. We thus feel that the whole subject warrants a fresh look in the light of recent findings.

 We would like to revisit the question of the gap induction by a magnetic field in a nanoscale $d$-wave superconductor.  While the general expectation that a magnetic field will induce additional components of the order parameter remains, the specific case of a small superconducting grain allows for sharp transitions between states with different orbital magnetic moment carried by pairs. These changes in magnetization can be observable in the case of small $d$-wave grains, as we will point out. 
Therefore qualitatively new effects can be expected in investigating small grains of $d$-wave superconductors. Earlier it was pointed out by Laughlin that the presence of nodes makes the $d$-wave superconducting states inherently unstable to the induction of novel components of the gap \cite{Laughlin98}. Similar effects of induction of additional components  are expected in the presence of a steady supercurrent \cite{Fogelstrom97, Volovik97}. The general source of instability of the pure $d$-wave gap is the presence of the nodes. 
A secondary component allows these nodes to be completely lifted. The most probable secondary components proposed in this context are $d+is$ \cite{Fogelstrom97} and $ d+id'$ \cite{Laughlin98} states. Both of these states will "seal the node of the gap". Both of the states break time reversal symmetry and thus will induce edge currents. The key difference between these two proposed states is that a $d+id'$ state is {\em chiral} and hence will have an orbital moment carried by Cooper pairs.

The appearance of an $id'$-wave competing order component is expected in the presence of an external magnetic field \cite{Laughlin98,Balatsky2000}, and has been used to explain some experiments \cite{Krishana97, Aubin99, Leibovitch08}. In the present paper we will consider values $H\sim H_{c2}$ for the external magnetic field, such that the high density of produced vortices almost destroys the superconducting state.  In this region of the parameter space, the action of the external field is able to induce a distortion of the original $d$-wave state into a $d_{x^2-y^2}+id_{xy}$ (briefly referred to as $d+id'$) bulk state, having an intrinsic angular momentum. That induces a movement across ground states with different symmetries, and consequent jumps in thermodynamic quantities.   
The low-energy quasiparticles sitting in the nodes of the $d$-wave superconductor (SC) have a vanishingly small gap and allow for the generation of a secondary component; this feature of $d$-wave SCs is called marginal stability, which is prevented in the $s$-wave SC by the presence of a finite gap on the Fermi surface \cite{Balatsky99}. The changes from the zero angular momentum state to a state with finite angular momentum will occur via a set of steps corresponding to $ L_z = 0, 2 \hbar, 4 \hbar,... $. 
These steps are small and are not observable in a bulk system since the number of Cooper pairs is large and the effect is therefore $ \propto 1/N_{pairs}$. Only in a sufficiently small grain, this staircase of angular momentum jumps becomes evident. In this paper we will analyze the angular momentum jumps in a $d$-wave grain as one sweeps the applied magnetic field. 
These jumps are similar to the Coulomb blockade charge jumps seen in quantum dots \cite{Kouwenhoven98}, see Table \ref{table1}. The underlying energy that controls these jumps is the energy of orbital moments in magnetic fields and thus is much smaller than the Coulomb energy controlling charge blockade phenomena. So in analogy with the Coulomb blockade we propose that the field-induced angular momentum changes in the Cooper pair states represent an {\em angular momentum} blockade that can still be seen in the susceptibility for a small superconducting grain. This is the main finding of this paper.

Existing results \cite{Gladilin04} allow us to exclude a similar blockade effect in ordinary $s$-wave superconductors, in agreement with a vanishing contribution of the angular momentum term to their free energy. Indeed, the magnetic susceptibility is therein proven to show a maximum for a well-defined value of the magnetic field  as soon as ultra small SC grains are considered.

We also point out that the effects discussed in this paper will equally be present in the $p$-wave superconductors. The obvious analogy will be the induction of $p+ip'$ superconductivity in the presence of a magnetic field due to the magnetic moment coupling to an external field. We will not specifically elaborate on that case but wish to point this obvious extension of the calculations presented.

The plan of this paper is as follows. First we will connect the field-induced component of the order parameter to the angular momentum. This will allow us to express the magnetic observables in terms of the angular momentum
itself, by adopting the Ginzburg-Landau description for the superconductor. We will conclude by analyzing the dependence of these observables on the external magnetic field, and by estimating the orders of magnitude involved in the effect. All the relevant numerical values and some technical details can be found in the Appendices.

\section{Angular momentum blockade}
 First we elaborate on the notion of angular momentum blockade. The low energy gaps induced by the magnetic field translate into very long length scales relevant for the formation of the new component of condensate. Therefore the formation of the induced gaps can be described by a continuum theory. In that limit the total angular momentum $L_z$ of the condensate is a good quantum number. Simple inspection shows (see below) that the pure $d$-wave state has zero angular momentum. The $d+id'$ state is a chiral state and has a finite angular momentum  $\neq 0$. Therefore any changes from the ground state with zero quantum number to a ground state with nonzero quantum number would proceed as a set of steps in $L_z$. The smallest steps of total angular momentum one can get in the SC grain will be in units of $2 \hbar$. Indeed we envision the set of incremental steps by which Cooper pairs convert from $d$ to $d+id'$ as a set of jumps in $L_z$. These jumps result in jumps of magnetization that can be seen only if the relative change is large. Hence it is expected that these changes will be seen only in small samples. This discussion is analogous to the Coulomb blockade that describes single electron charge jumps, Table\ \ref{table1}.

\begin{figure}
\centering
\subfigure [ \;Competing order]
{\includegraphics[width=6cm]{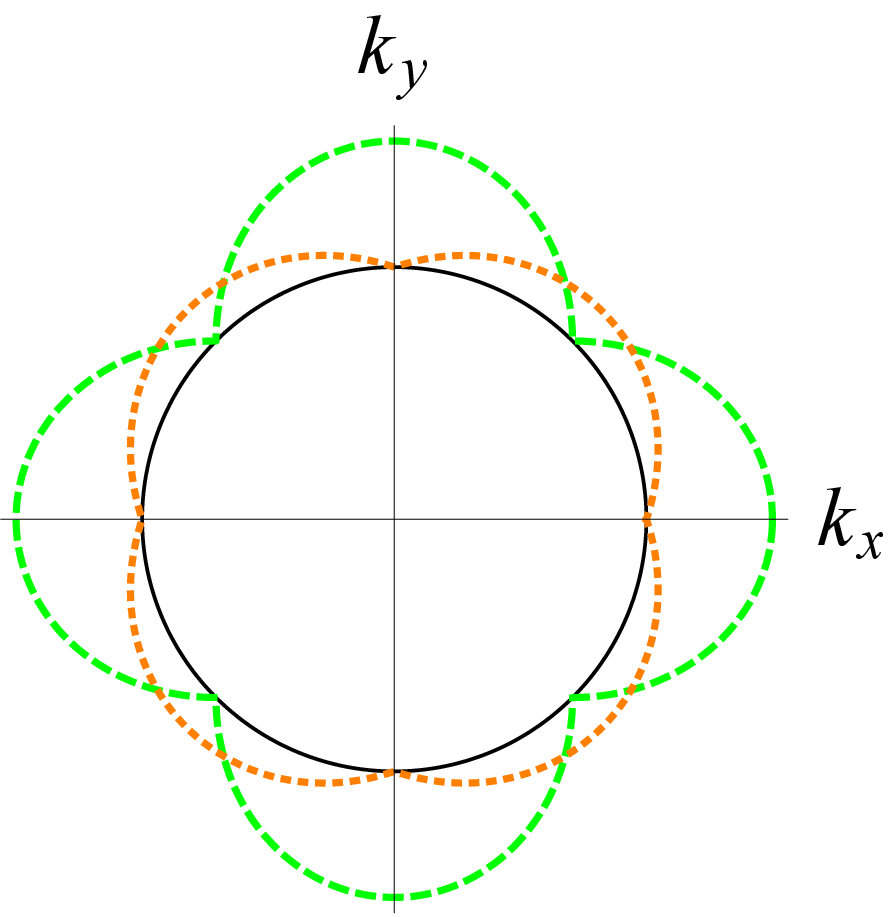}} \quad
\subfigure [ \;Full gap]
{\includegraphics[width=6cm]{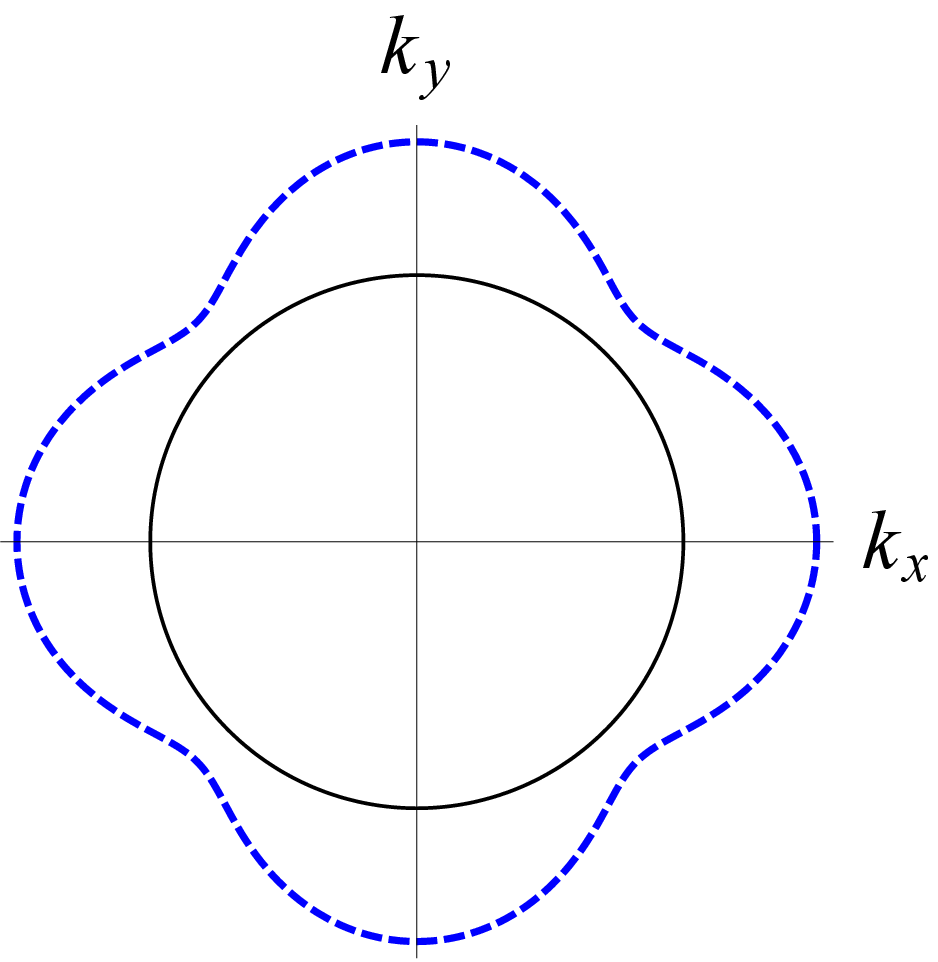}}
\caption{
(a) The black circle representing the Fermi surface is shown together with the angular dependent gap amplitudes.  The dashed green curve is the $d$-wave gap $\Delta_0$, and the smaller dotted orange curve is the induced $d'$-wave gap $\Delta_1$. For display purposes, the relative proportion of the gaps has not been preserved in the plot. \,(b) Outside the black circular Fermi surface, the total gap $\sqrt{\vert\Delta_0\vert^2+\vert\Delta_1\vert^2}$ of the fully gapped $d+id'$ state is shown in blue. The former gapless node lines of the $d$-state, which are the diagonals w.r.t. the $[100]$ and $[010]$ crystallographic directions, are sealed by the induced imaginary component $id'$, which renders the overall $d+id'$ state fully gapped.}
\label{gaps}
\end{figure}

To illustrate the mechanism we will use the Ginzburg-Landau theory for an high-$T_c$ SC in an external magnetic field, which is written in terms of the complex order parameter

\be \Delta(\vec{r},\theta)=\exp[i\Phi(\vec{r})]\Delta(\theta)\;,
\ee
where $\Phi(\vec{r})$ is the phase of the order parameter, and $\theta$ refers to the dependence of the pair wave function on the position on the (2D) Fermi surface. The physical origin for the generation of the second component $id'$ resides in the induced bulk magnetic moment $\langle M_z\rangle$, and the relevant interaction is the $\langle M_z\rangle B$ coupling to the magnetic field $B \parallel z$, where $B$ is the magnetic induction in presence of an external field $H$. The $d$-wave state before turning on the magnetic field can be regarded as a superposition of the pairs with orbital momentum $L_z=\pm 2\hbar$:
\be
\Delta_0(\theta)=\Delta_0\,\cos 2\theta=\frac{\Delta_0}{2}\left[\exp(2i\theta)+\exp(-2i\theta)\right]\;,
\ee
where $\Delta_0$ is the gap magnitude of the $d$ component, and a 2D geometry of the Fermi surface is considered in the model, because of the layered structure of the cuprates. The external field $H$ has the effect of shifting these two $L_z=\pm 2\hbar$ components, linearly and with opposite signs as:
\be
\Delta_0(\theta)\rightarrow \frac{\Delta_0}{2}\left[(1+\eta B)\exp(2i\theta)+ (1-\eta B)\exp(-2i\theta)\right]=\Delta_0(\theta)+i B \vert\Delta_1(\theta)\vert\;,
\ee
where $\vert\Delta_1(\theta)\vert\propto \eta \sin 2\theta$ is the modulus of the $d'$ component and $\eta$ is the coupling constant. Unlike each of its two components taken alone (having nodes of the gap), the full $d+id'$ state is fully gapped \cite{Balatsky2000, Balatsky97}, see Fig.\ \ref{gaps}.

To a good approximation, the angular momentum $L_z$ of the whole sample about $z$ is related to $\Delta_0,\Delta_1$ (see Appendix \ref{appendangularmomentum} for derivation) by:

\be \vert\Delta_1\vert=\vert\Delta_0\vert \frac{\vert\langle L_z\rangle \vert}{ \,2 \hbar\; g\;(V/a^2 c)}\;.\ee

Here the components of the order parameter have been assumed to be spatially uniform within the sample, which is justified to a good approximation for sample sizes much smaller than the characteristic length of variation of the SC wave function;
$(a^2 c)$ denotes the unit cell volume of the SC's crystal structure (being $c$ the edge $\parallel \hat{z}\parallel B$ field, $a \approx 4 $\AA, $c \approx 10 $\AA ), $V$ the volume of the SC sample, and $g\sim 0.1$ the superconducting fraction of the material. \\

\begin{table}
    \begin{tabular}{ | p{8.5cm} |p{8.5cm} |}
   \hline
    \textbf{COULOMB BLOCKADE} & \textbf{ANGULAR MOMENTUM BLOCKADE}  \\  \hline \hline
    Charge $Q$  = deposit of electric potential energy &  Angular momentum $L$ = deposit of mechanical energy  \\ \hline
    Electrostatic potential $V_{\text{max}}$: beyond which there is spontaneous discharge & Angular velocity  $\omega_{\text{max}}$: beyond which the centrifugal stress results in breaking\\ \hline
    Discharge:  decreasing $V$  &   Discharge:  decreasing $\omega$  \\ \hline
   Capacitance  $C\, \propto$ area of the plates  & Moment of intertia $I_{\text{in}}$ : increasing with spreading of the mass distribution about the rotation axis \\  \hline

   $Q=C\;V$  &  $L= I_{\text{in}}\;\omega$ \\  \hline

    Energy $E=CV^2/2$  &  Energy $E=I_{\text{in}}\; \omega^2/2$ \\  \hline
Current $i$  & Torque $\tau$ \\  \hline
    \end{tabular}

\caption{Mapping of the fundamental quantities and relations relevant to the Coulomb blockade to their corresponding quantities for the angular momentum blockade.}
\label{table1}
\end{table}

Let us define the non-negative integer variable $l\equiv |\langle L_z \rangle|/(2 \hbar)$ and the characteristic area $S_0\equiv a^2 c/(d\cdot g)$, where  $d$ is the sample thickness and $V=A d$ is the volume. This gives

\be \label{blockade}  \vert\Delta_1\vert=l\;\vert\Delta_0\vert\frac{S_0}{A} \;.\ee

In the case of an isotropic gradient tensor $K_{ij}$, and if the grain is small enough to allow us neglect the spatial dependence of $\Delta_0(\vec{r}), \Delta_1(\vec{r})$, the GL functional takes the form
\be  F=\int_{V=Ad}{d^3r \; \left[\alpha \,\vert \Delta_0\vert^2+\frac{\beta_0}{4}\vert\Delta_0\vert^4+K\left\vert(-i\frac{e}{c}\vec{A})\Delta_0\right\vert^2-\eta B |\Delta_0||\Delta_1|+\frac{N_0}{2}\vert\Delta_1\vert^2\right]}\;,\ee
provided that we omit the magnetic field self-energy term, which does not affect the behavior of the magnetization or the susceptibility that we are going to describe.
In the Coulomb gauge $\vec{A}(\vec{r})=\frac{B}{2}(-y,x)$, Eq.\ (\ref{blockade}) gives
\be F(B,\langle L_z \rangle)=\int_{V=Ad}{d^3r \; \left[\alpha \,\vert \Delta_0\vert^2+\frac{\beta_0}{4}\vert\Delta_0\vert^4+K\frac{e^2}{c^2}\frac{B^2}{4}r^2\left\vert\Delta_0\right\vert^2-\eta B \;|\Delta_0|^2\frac{S_0}{A}\,l+\frac{N_0}{2}\vert \Delta_0\vert^2 \frac{S_0^2}{A^2}\,l^2\right]}\;.\ee

The coupling constant $\eta$ is derived and discussed in the Appendix of Ref.\ \cite{Balatsky2000}.

In the same way as the Coulomb blockade leads to jumps in charges and in charging energy, the analogous phenomena for the angular momentum blockade involve changes in angular momentum and magnetization. There will be jumps and spikes in magnetization and susceptibility, respectively, which we will now consider.  The free energy $F$ is minimized by an optimal integer value of $l$ for each value of the applied field $B$.  Hence $l$ forms a stepwise function of $B$ with steps at the switching fields $B=B_l$.  The derivative $\partial l/\partial B$ is a sequence of delta function spikes at the switching fields
%$$
%\partial l/\partial B \sim \sum_l \delta(B-B_l)\;,
%$$ 
that should be experimentally measurable.  We will now consider these phenomena in detail.
%, in the form of $\left(\eta |\Delta_0|^2 S_0/A\right)$-sized jumps in the magnetization, and corresponding delta-like spikes in %the susceptibility, as the external magnetic field is swept. }

%A numerical estimate for the electronic density of states $N_0$ comes from the comparison between
%\be n_{(particle \;density)}=\frac{p_F^3}{3\pi^2 \hbar^3} \quad \text{(for electrons)};
%\ee
%and the electronic density of states at the Fermi surface 
%\be 
%N_0=\frac{m_e p_F}{2\pi^2 \hbar^3}=\;\cdots\; = \left(\frac{3}{\pi}\right)^{1/3} \frac{m_e \, n^{1/3}}{2\pi \hbar^2}\,, 
%\ee
%whence the unit cell dimension ($4$\AA $\times 4 $\AA $\times 10 $\AA) produces the order of magnitude 
%\be 
%N_0 \approx 2.4 \times 10^{33} \text{ erg}^{-1} \text{cm}^{-3}\;.
%\ee

The free energy density can be rewritten as 
\be 
F_v(|\langle L_z \rangle|=2 \hbar l, B)\equiv \frac{F(|\langle L_z \rangle|=2 \hbar l, B)}{V}=const+a B^2-b B l+ c l^2\;,
\ee
forming a piecewise parabolic function, where the following constants were introduced for convenience:
$a\equiv K\frac{e^2}{c^2}\frac {A\vert \Delta_0 \vert^2}{8\pi},
b\equiv \eta |\Delta_0|^2 \frac{S_0}{A},
c\equiv \frac{N_0}{2} |\Delta_0|^2 \frac{S_0^2}{A^2}$.
The critical values $B_{n}$ where the optimal $l$ switches from $l=n$ to $l=n+1$
are obtained from the level crossings of the functions $F(l=n, B)$ and $F(l=n+1, B)$, which gives
\be 
B_{n}=\frac{c}{b}(2n+1)\;.
\label{criticalfields}
\ee
Hence the angular momentum value $\langle L_z \rangle=n\,(2\hbar)$ is attained for $B_{n-1} < B < B_{n}$.
The integer $n$ will henceforth denote the integer that minimizes $F_v$, which is given by
the integer part of $(1+(b/c) B)/2$.
%, 
%which means that $\langle L_z \rangle=n \,(2\hbar)$ for any value of the field in $\frac{c}{b}(2n-1)<B<\frac{c}{b}(2n+1)$.
The magnetization and the susceptibility then become
%Then, for $B \in \frac{c}{b}\times (2n-1,2n+1) $ it stands 
\be 
M_v=-\frac{\partial F_v}{\partial B}= -2 a B +b n  \;,
\ee
and
\be 
\chi_v=\frac{\partial M_v}{\partial B}=-2a+b \sum_n \delta(B-B_n) \;.
\ee

%Now for $B\in \frac{c}{b}\times (2n-1,2n+1)$ it is $n=\left[ \frac{1+\frac{b}{c}B}{2} \right]$, where here and henceforth the notation %$\left[ \cdots \right]$ denotes the arithmetic \textit{floor function}, so 
% \be 
% F_v= const+a B^2-b B \left[ \frac{1+\frac{b}{c}B}{2} \right]+ c \left[ \frac{1+\frac{b}{c}B}{2} \right]^2\;\;. 
% \ee
% \be 
% M_v=-\frac{\partial F_v}{\partial B}= -2 a B +b\; \left[ \frac{1+\frac{b}{c}B}{2} \right]\;\;. 
% \ee
% \be 
% \chi_v=\frac{\partial M_v}{\partial B}=-2a+b \sum_n \delta(B-B_n) %  \quad (\text{\textbf{dimensionless}})
% \ee

Except for the non-regular behavior at critical values (\ref{criticalfields}) of the field, the magnetic susceptibility should have the constant value $\chi=-\frac{1}{4\pi} K \frac{e^2}{c^2} A \vert\Delta_0\vert^2$. The free energy density has cusps joining different parabolic arcs at each critical value for the magnetic field. The magnetization has constant jumps inversely proportional to $V$ (connecting regions of linear behavior). Finally, the susceptibility has $\delta$-like spikes inversely proportional to $V$.

\section{Behavior of magnetic observables}

Let $x\equiv \frac{A}{S_0}$ be the dimensionless area and $y\equiv \frac{b}{c} B$ 
the dimensionless magnetic field.  In terms of these quantities we have
$a=\frac{1}{8\pi}K\frac{e^2}{c^2} S_0 \vert \Delta_0 \vert^2 x,
b= \eta |\Delta_0|^2 \frac{1}{x},
c= \frac{N_0}{2} |\Delta_0|^2 \frac{1}{x^2},
%n=\left[\frac{1+y}{2}\right],  %% NOT NEEDED
B=\frac{c}{b}y=\frac{N_0}{2\eta}\frac{y}{x}\approx \frac{E_{\rm F}}{2\mu_{\rm B}}\frac{y}{x}\;\,.$
Then the free energy density, magnetization, and susceptibility are given by
\be 
F_v = const+\frac{1}{8\pi}K\frac{e^2}{c^2} S_0 \vert \Delta_0 \vert^2   
\left(\frac{E_{\rm F}}{2\mu_{\rm B}}\right)^2\frac{y^2}{x}-\frac{N_0|\Delta_0|^2}{2}\frac{yn-n^2}{x^2} \;,
\ee
\be
M_v= -\frac{1}{8\pi}K\frac{e^2}{c^2} S_0 \vert \Delta_0 \vert^2  
\frac{E_{\rm F}}{\mu_{\rm B}} \; y +\frac{N_0 |\Delta_0|^2 \mu_{\rm B}}{E_{\rm F}} \frac{n}{x} \;,
\label{magnetizationdimensionless}
\ee
\be
\chi_v =-\frac{1}{4\pi}K\frac{e^2}{c^2} S_0 \vert \Delta_0 \vert^2 x +\frac{2\eta^2 |\Delta_0|^2}{N_0} \sum_n \delta(y-(2n+1))\;. 
\label{susceptibilitydimensionless}
%\quad (\text{\textbf{dimensionless}})
\ee

The magnetization (\ref{magnetizationdimensionless}) in terms of the original dimensionful parameters takes the form
\be
\quad M_v(B)= -\frac{1}{8\pi}K\frac{e^2}{c^2} \frac{(a^2 c)}{g \cdot d} \vert \Delta_0 \vert^2  \frac{E_{\rm F}}{\mu_{\rm B}} \; y +\frac{N_0 |\Delta_0|^2 \mu_{\rm B}}{E_{\rm F}} \frac{a^2c}{g  A d} n \;.
\label{magnetizationdimensionful}
\ee
The second term is responsible for jumps of $M_v$ at each critical value of the magnetic field.  All these jumps have the same amplitude and mutual spacing.  The first term is due to the minimal coupling between the magnetic gauge potential and the order parameter.  This contains the gradient tensor, and leads to a smooth monotonic variation of the magnetization between consecutive jumps.

The magnitude of the magnetization jumps $\Delta M_{\rm blockade}$ associated to the blockade mechanism increases if the sample area $A$ decreases.  
%The smaller is the area $A$ of the SC sample (orthogonally offered to the external field), the higher is the importance of the 
%jumps $\Delta M_{\rm blockade}$ associated to the blockade mechanism, and their evidence in the pattern of $M_v(B)$ for a given
%thickness $d$. 
In fact, the jump size grows as $A^{-1}$ although remaining subleading with respect to the continuous variation 
$\Delta M_{\text{cont}}$ between jumps, even for tiny areas $A$. 
The first term in Eq.\ (\ref{magnetizationdimensionful}) is unaffected by $A$.

\begin{figure}
\vspace{0.4cm}
\centerline{\scalebox{0.7}{\includegraphics{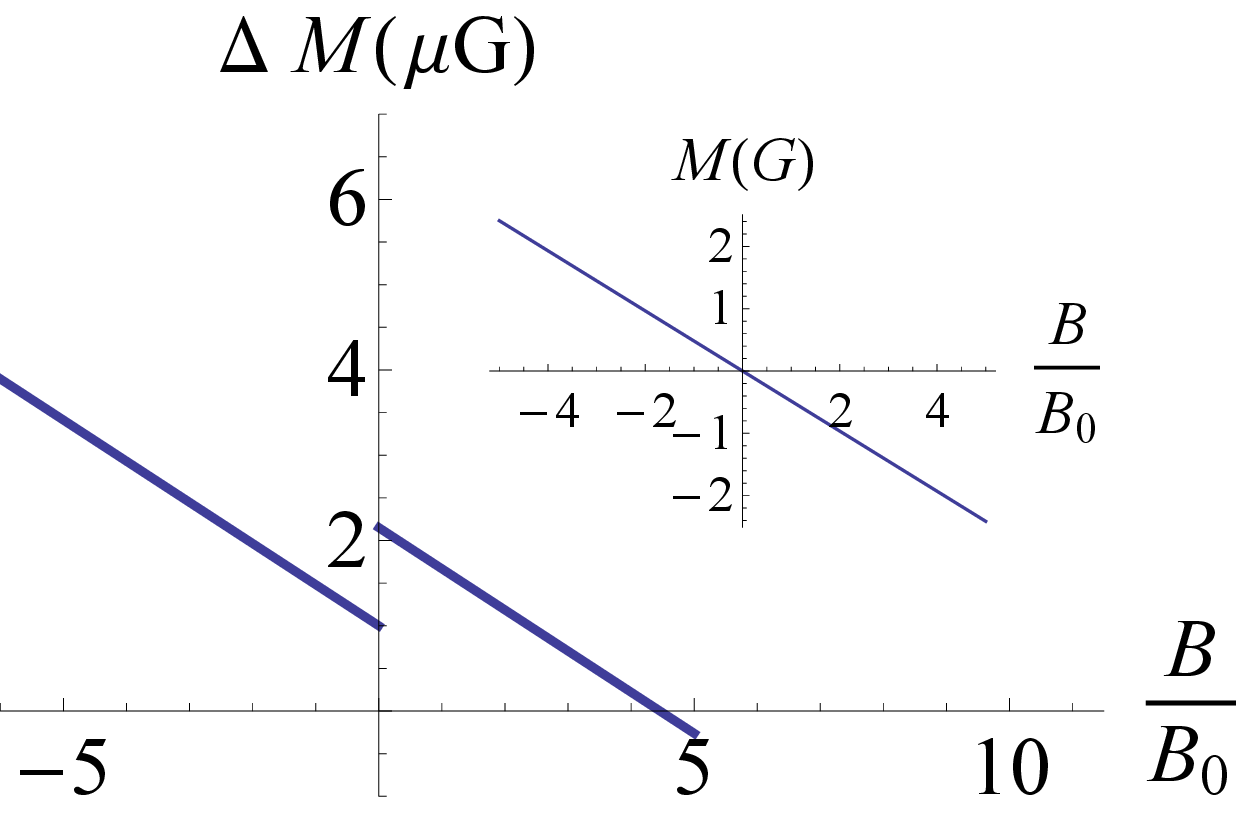}}}
%\vspace{0.5cm}
\caption{Magnetization $\Delta M$ of the sample (up to an additive constant), plotted in microgauss for the example case of area $A=(100 \text{ nm})^2$ and thickness $d=100 \text{ \AA}$, as a function of the magnetic field $B/B_0$, where $B_0\equiv N_0(a^2c)/(2\,\eta \,g \,A\,d)\sim 0.7 \text{ tesla}$ for such a grain volume. In this plot, the field variable $B/B_0$ is centered about a critical level crossing point (i.e. any odd integer) and then zoomed by a $10^6$ factor.  It is evident how this sample volume entails magnetization jumps $ \sim 1\mu \text{G}$\;.
The background gradient tensor contribution to the magnetization is by far the leading one, being the jumps comparatively \textit{micro}-scopic (see formula (\ref{jumpsratio})): it is represented in units of gauss (with no scaling for the axis) in the inset at the top right, for a sample of thickness $d=100 \text{ \AA}$\,.}
\label{withinset}
\end{figure}

\begin{figure}
\vspace{0.4cm}
\centerline{\scalebox{0.7}{\includegraphics{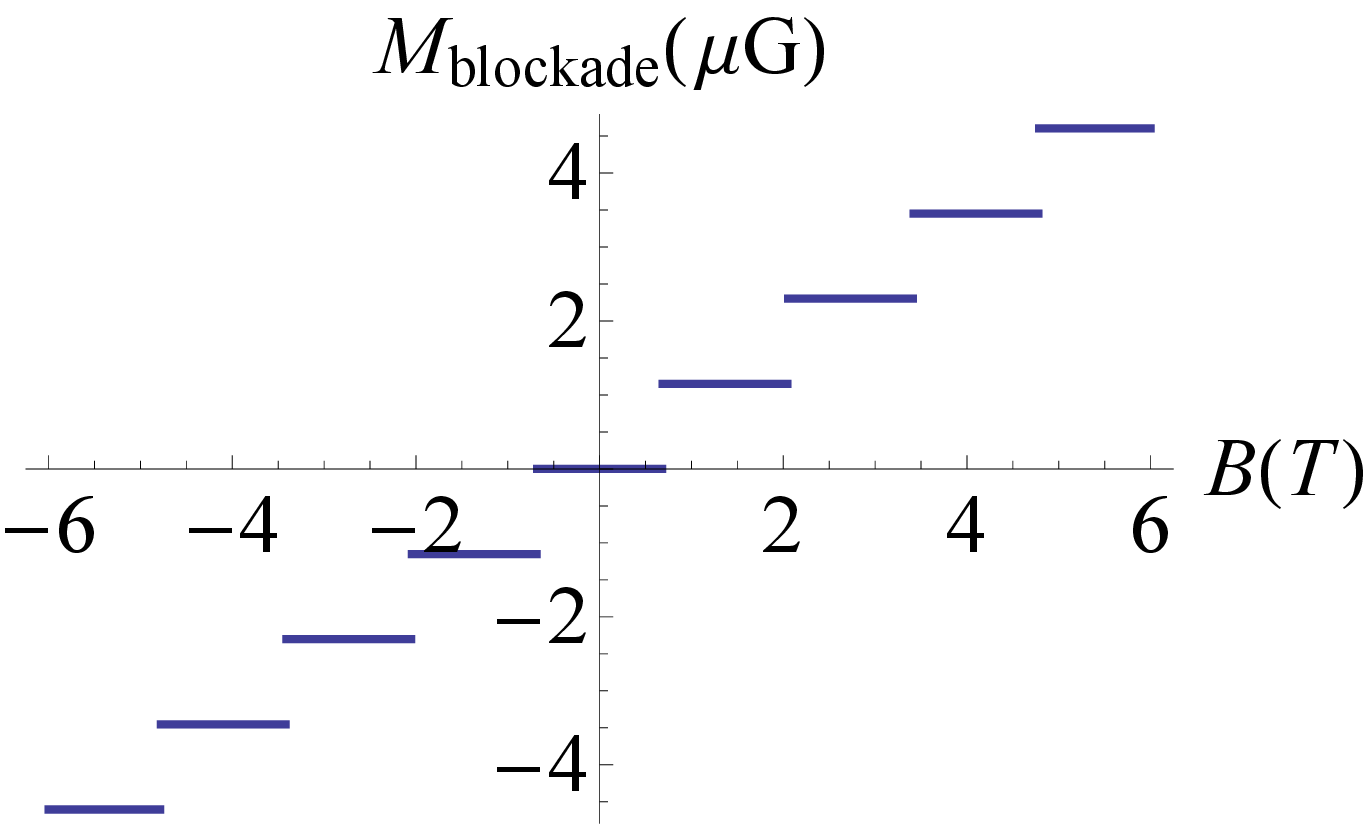}}}
%\vspace{0.5cm}
\caption{Blockade contribution to the magnetization of the sample, as a function of the magnetic field $B$ expressed in tesla, for a cylindrical grain of thickness $d=100$ {\AA} and area $A=(100 \text{ nm})^2$.   The system displays a staircase of magnetization jumps of the order of $\sim 1 \mu$ G, which are mutually spaced by regular intervals $\Delta B \sim 1.4$ T.}
\label{blockadeonly}
\end{figure}

\begin{figure}
\vspace{0.4cm}
\centerline{\scalebox{0.55}{\includegraphics{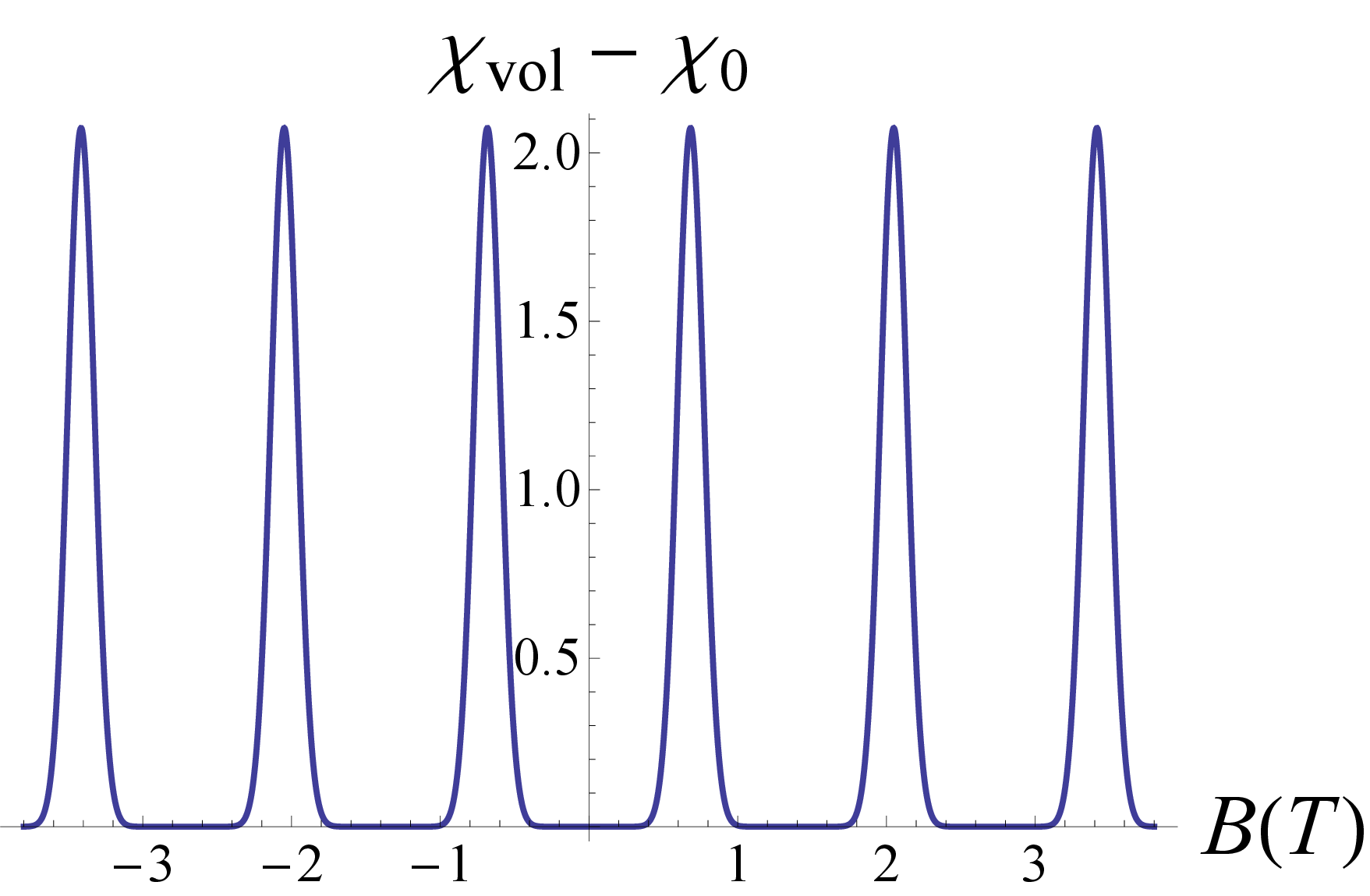}}}
%\vspace{0.5cm}
\caption{(Dimensionless) volume susceptibility of the sample, plotted againts the  magnetic field $B$ for a cylindrical sample of thickness $d=100 \text{\AA}$ and area $A=(100 \text{ nm})^2$; the vertical axis is in units of $10^{-9}$. There is a background smooth contribution $\chi_0$ to the susceptibility (about $-3 \cdot 10^{-4}$ for this grain) due to the gradient term of the GL functional, which is proportional to the area $A$ of the sample, and $\delta$-like spikes above it, due to the angular momentum blockade, which are $\propto\,V^{-1}$, and mutually spaced by regular intervals $\Delta B\sim $ 1.4 \text{ T}\;.}
\label{plotsusceptibility}
\end{figure}

The ratio between the two contributions to $M_v(B)$ is
\be
\frac{\left\vert\Delta M_{\rm blockade}\right\vert}{\left\vert\Delta M_{\rm cont}\right\vert}=\frac{A_{\rm cr}}{A}\;, 
\label{jumpsratio}
\ee 
where $A_{\rm cr}\equiv \frac{4 \pi N_0}{K}\left(\frac{c\,\mu_{\rm B}}{e E_{\rm F}}\right)^2\sim 1.20 \; \text{\AA}^2$ is a \textit{critical area} independent of any macroscopic feature of the SC sample. The magnetization and susceptibility given by Eqs.\ (\ref{magnetizationdimensionless})-(\ref{susceptibilitydimensionless}) are plotted in Figs.\ \ref{withinset}-\ref{plotsusceptibility}.  In the plots a cylindrical sample is assumed with dimensions $V=d \cdot A\equiv 100 \text{ \AA}\times (100 \text{ nm})^2$. The blockade contribution alone to the magnetization is depicted in Fig.\ \ref{blockadeonly}.  The dimensionless term $K\frac{e^2}{c^2} S_0 \vert \Delta_0 \vert^2 x$ is proportional to the area $A$, this is the only parameter on which it depends (among the parameters thickness $d$, area $A$ and magnetic field $B$), and its value affects the plot of the volume susceptibility in Fig. \ref{plotsusceptibility}. 

\section{Conclusions}

In conclusion we propose the notion of {\em angular momentum blockade} in granular $d$-wave superconductors. The effect is due to abrupt changes in the angular momentum of the electron liquid in the condensate. The pure $d$-wave state has $L_z = 0$. As an out-of-plane magnetic field is applied to the sample, a chiral $d'$-component  is induced. The angular momentum of Cooper pairs has to vary accordingly. This process of conversion of momentum, from $L_z = 0$ to a finite value given by the induced $d'$ component, proceeds in a series of steps where individual Cooper pairs acquire finite angular momentum. These steps are what we call angular momentum blockade. Jumps in the angular momentum of the condensate result in jumps of the magnetization, which we calculate. We find that these steps occur at well-defined values of magnetic field.  We also have defined a "characteristic area" $S_0$ ($\propto$ inverse thickness $d^{-1}$), providing a scale for the change of induced gaps $\vert\Delta_1\vert$ between two critical fields causing steps in magnetization, in units of the $d$-wave gap amplitude $\vert\Delta_0\vert$. On the other hand, we have provided a "critical area" $A_{\text{cr}}$,
which is the area scale related to the relevant weight between the two distinct contributions entering the magnetization variation (within the same magnetic field range considered above): the blockade contribution w.r.t. the gradient tensor one. \\
As illustrated in Figs. \ref{withinset},\ref{blockadeonly}, a nanoscale grain of typical volume $10^5\,\text{nm}^3$ should exhibit significant magnetization jumps of the order of 1 microgauss, which are encountered at critical magnetic fields whose uniform mutual spacings we estimate being of the order of $\sim$ 1 tesla. These results suggest that sensitive magnetization experiments might be able to see these jumps. 
Finally in Fig.\ \ref{plotsusceptibility} two different contributions to the magnetic susceptibility are highlighted: the gradient term contribution, which is a constant background value proportional to the area $A$ of the sample, and the angular momentum blockade contribution, consisting of $\delta$-like spikes $\propto\,V^{-1}$, and mutually spaced by regular intervals $\Delta B$ ($\sim $ 1 tesla for the example grain). \\

\textit{Acknowledgements}  This work has been supported by the Swedish Research Council grants VR 621-2012-298, VR 621-2012-3984, ERC and DOE. We are grateful to B. Altshuler for an earlier discussion  of angular momentum blockade.

\appendix 
\section{Relation between the angular momentum and the gaps for the $d$ and the $d'$ components}
\label{appendangularmomentum}

We refer to the components of the SC wave function, having $d_{x^2-y^2}$ and $d_{xy}$ symmetry respectively, as $\psi_0$ and $\psi_1$.
The angular momentum becomes
\be
\langle L_z\rangle=\langle \psi_0(\theta)+ \psi_1(\theta) \vert -i \hbar \partial_\theta  \vert \psi_0(\theta)+\psi_1(\theta) \rangle=
\nonumber
\ee
\be
= -i \hbar \int{dV \,\frac{d \theta}{2\pi} 2[\psi_0^* \psi_1 \cos(2 \theta)^2 - \psi_0 \psi_1^* \sin(2 \theta)^2]} = -i\hbar\int \,dV\; [\psi_0^* \psi_1 - \psi_0 \psi_1^*] \; . 
\ee
In terms of  respective moduli and phases $\psi_0 = |\psi_0| \exp(i\nu_0)$, $\psi_1 = |\psi_1| \exp(i\nu_1)$, the  above equation becomes
\be
\langle L_z\rangle  = \int dV {[-\hbar 2|\psi_0||\psi_1|\sin(\nu_0 - \nu_1)]} \; .
\ee
Because of the global U(1) symmetry we can arbitrarily choose $\nu_0 = 0$ for this discussion, and the free energy is minimized when the relative phase $(\nu_0 - \nu_1) = \pm \pi/2$.
For a spatially uniform wave function, i.e., a spatially uniform order parameter, it is
\be
\langle L_z\rangle  = V {[-\hbar 2|\psi_0||\psi_1|\sin(\nu_0 - \nu_1)]} \; .
\ee
For the minimizing relative phase $\pm \pi/2$ this gives
\be
\langle L_z\rangle  = V \left[{-\hbar 2|\psi_0|^2\frac{|\Delta_1|}{|\Delta_0|}}\right] \; ,
\ee
since $\vert\Delta_1\vert / \vert \Delta_0\vert = \vert\psi_1\vert / \vert\psi_0\vert$.
With $a^2c$ the unit cell volume, 
$d$ the sample thickness, and $V=Ad$ the sample volume,
the number of Cooper pairs  $N_{\text{sc}}$ in the grain
is approximately
\be
V|\psi_0|^2=N_{\text{sc}}=g\,\frac{V}{a^2 c} \; ,
\ee 
where $g\sim 0.1$ is the high-$T_c$ superconducting fraction.  This gives the estimate
\be
|\Delta_1|=|\Delta_0| \frac{|\langle L_z\rangle|}{2 \hbar g V/(a^2 c)} \; .
\ee

\section{Numerical values}
For allowing better checks, numerical values listed below come with more digits than the significative ones. \\
$$\begin{array}{lr}
\text{Edges of the SC unit cell} & a \sim 4 \text{\AA}, \; c \sim 10 \text{\AA} \\
\text{Coherence length} & \xi \sim 10 \text{\AA} = 10^{-7} 	\text{cm} \\
\text{Superconducting fraction} & g \sim 10^{-1}\\
\text{Unperturbed } d_{x^2-y^2}\text{-wave gap} & \vert\Delta_0\vert \sim 50 \text{ meV}=8.011\times 10^{-14} \text{ erg} \\
\text{Density of states} & N_0 \sim 2.4\times 10^{33} \text{ erg}^{-1}\text{ cm}^{-3}\\
\text{Fermi energy } & E_F \sim 1.98\times 10^{-12} \text{ erg}\\
\text{Thickness } & d\equiv 100 \text{ \AA} =10^{-6} \text{cm}\\
\text{"Characteristic area"}& S_0\equiv\frac{(a^2c)}{d\cdot g} \sim 1.6\times 10^{-15}\text{cm}^2= 16\text{ \AA}^2 \\
\text{Modulus of the gradient tensor} & K \sim \hbar^{-2}\xi^2 N_0 \sim 2.156 \times 10^{73} \text{ cm}^{-1} \text{s}^{-2} \text{erg}^{-3} \\
\text{Dimensionless recurrent combination} & K\frac{e^2}{c^2}S_0|\Delta_0|^2 \sim 5.68\times 10^{-8}\\
\text{"Critical area"} & A_{\text{cr}}\sim 1.20 \times 10^{-16} \text{cm}^2=1.20\;\text{\AA}^2\\
\end{array}$$\\

\end{document}